\documentclass[%
 reprint,
 superscriptaddress,
 amsmath,amssymb,
 aps,
]{revtex4-1}

\usepackage{graphicx}
\usepackage{dcolumn}
\usepackage{bm}
\usepackage{xcolor}

\begin{document}


\title{The Mass of Short-Range Correlated Nucleon}

\author{Rong Wang}
\email{rwang@impcas.ac.cn}
\affiliation{Institute of Modern Physics, Chinese Academy of Sciences, Lanzhou 730000, China}
\affiliation{University of Chinese Academy of Sciences, Beijing 100049, China}

\author{Xurong Chen}
\email{xchen@impcas.ac.cn}
\affiliation{Institute of Modern Physics, Chinese Academy of Sciences, Lanzhou 730000, China}
\affiliation{Institute of Quantum Matter, South China Normal University, Guangzhou 510006, China}

\author{Taofeng Wang}
\email{tfwang@buaa.edu.cn (corresponding author)}
\affiliation{School of Physics, Beihang University, Beijing 100191, China}


\date{\today}

\begin{abstract}
The nucleon properties and structure should be modified by short-range correlations (SRC) among nucleons.
By analyzing SRC ratio data, we extract the mass of nucleon in the SRC pair
and the expected number of pn-SRC pairs in deuterium,
under the assumption that the SRC nucleon mass is universal in different nuclei.
The nucleon mass of a two-nucleon SRC pair is $m_{\text{SRC}}=852\pm 18$ MeV,
and the number of pn-SRC pairs in deuterium is $n^{\text{d}}_{\text{SRC}}=0.021\pm 0.005$.
The mass deficit of the strongly overlapping nucleon can be explained by the
trace anomaly contribution to the mass in QCD or alternatively by the vacuum energy in the MIT bag model.
\end{abstract}

\pacs{24.85.+p, 13.60.Hb, 13.85.Qk}
\maketitle


Nucleon-nucleon short-range correlation (NN-SRC) is one typical microscopic structure of the nucleus,
resulting from a strong overlap of the wave functions of two nucleons.
NN-SRC has attracted a lot of experimental interests in the last decades,
as it is an interesting, new phenomenon beyond the description of mean field theory \cite{Hen:2016kwk,Atti:2015eda,Arrington:2011xs,Frankfurt:1988nt},
which has furthered our understanding of many-body problems and high density cold nuclear matter.
The nucleons in NN-SRC are strongly interacting at a very short distance
that is comparable with the nucleon radius. Kinematically, the two nucleons inside a SRC pair have large relative momenta,
however the center-of-mass (c.m.) momentum of the pair is small, of the value of Fermi momentum
\cite{Aclander:1999fd,Subedi:2008zz,Duer:2018sby,Cohen:2018gzh}.
The nucleons in the SRC pair have high momenta while the mass of the SRC pair is limited,
hence the nucleons in SRC are of high virtualities.
Recently, the high momentum distribution of the SRC nucleon is explained successfully under
a generalized contact formalism (GCF) arisen from the universal contact term of Fermion interaction at high momentum
\cite{Weiss:2015mba,Weiss:2016obx,Weiss:2020mns}.
Moreover, SRC pairing has some influences on the nucleon structure and the nucleon properties,
for the nucleons in SRC pair interact strongly.

Experimental evidences for SRC come from exclusive \cite{Aclander:1999fd,Subedi:2008zz,Duer:2018sby,Cohen:2018gzh,Tang:2002ww,Piasetzky:2006ai,Hen:2014nza}
and inclusive measurements \cite{Frankfurt:1993sp,Egiyan:2005hs,Fomin:2011ng,Schmookler:2019nvf}
of quasi-elastic scattering between a probe and a SRC pair, showing back-to-back
high momentum motions of two final nucleons.
The exclusive measurements find that about 90\% NN-SRC pairs are proton-neutron pairs
\cite{Piasetzky:2006ai,Subedi:2008zz,Hen:2014nza,Duer:2018sxh},
which may explain the difference between the proton momentum distribution
and the neutron momentum distribution in a proton-neutron imbalanced system.
The pn-pair dominance in the SRC configurations indicates that the SRC pair
is induced by the strong tensor force \cite{Schiavilla:2006xx,Alvioli:2007zz,Neff:2015xda}.

Nucleon mass is a fundamental question in particle and nuclear physics.
But its origins are still under discussions \cite{Ji:1994av,Ji:1995sv,Lorce:2017xzd},
with few experimental tests. As the nucleon is not an elementary particle,
its bulk properties should be derived from the dynamics of its constituents.
In the Standard Model, the underlying theory describing the nucleon
is quantum chromodynamics (QCD). The extraordinary features of QCD
-- confinement and asymptotic freedom -- make the exact solution of a system
of hadron size unavailable. Nonetheless, with the development of computing technology,
lattice QCD (LQCD) has achieved some interesting numerical results on nucleon mass decomposition
\cite{Yang:2017erf,Yang:2018nqn}, in spite of providing few insights into the underlying physics.
According to the analysis of QCD energy-momentum tensor, the nucleon mass is
composed of four terms: the quark mass term $M_{\rm m}$, the quark energy term $M_{\rm q}$,
the gluon energy term $M_{\rm g}$, and the trace anomaly term $M_{\rm a}$ \cite{Ji:1994av,Ji:1995sv}.
The trace anomaly only appears in the renormalization of loop diagrams.
In MIT bag model, there is a clear physical picture.
The nucleon mass is attributed to the internal energies
of the massless particles (quarks and gluons) and the vacuum energy of the ``bag" volume \cite{Chodos:1974pn,Chodos:1974je}.
The conceptually simple assumption of the MIT bag model is that the fields are confined
to a finite space, which actually captures the confinement property of QCD.

On the experimental side, it is essential to look for the measuring method
and the choice of data which is sensitive to the nucleon mass structure.
The nucleus is made of nucleons and so
naturally provides a good platform to study the nucleonic properties.
The interaction between nucleons can be exploited to probe quark and gluon dynamics.
Nowadays, electron-ion colliders are widely discussed worldwide \cite{Accardi:2012qut,AbelleiraFernandez:2012cc,Chen:2018wyz,Chen:2020ijn}.
They will resolve several mysteries on nuclear effects and nucleon structure.
To determine the mass of a nucleon which interacts strongly with a neighboring nucleon
may shed some lights on the origin of the nucleon mass and the characteristics of nonperturbative QCD.
In this letter, we extract the nucleon mass in NN-SRC as well as the number
of SRC pairs in a deuteron, based on the inclusive measurements of high energy electron-nucleus scattering.

To search for NN-SRC using high energy electron-nucleus scattering, one measures
the nucleons at momenta larger than the Fermi momentum.
In electron-nucleus interaction, a virtual photon is exchanged between the electron
and the nucleus, and the Bjorken variable $x_{\rm B}=Q^2/(2m_{\rm N}\nu)$ is determined by the photon kinematics.
$Q^2$ and $\nu$ are the minus of the photon's four momentum square and the energy of the photon
in the nucleus' rest frame, respectively. For the definition of the Bjorken variable in experiment,
$m_{\rm N}=0.938$ MeV is taken to be the mass of the free nucleon.
Usually, the requirement $1.5<x_{\rm B}<2$ is applied to select
the quasi-elastic scattering of the electron on the NN-SRC pair \cite{Frankfurt:1993sp,Egiyan:2005hs,Fomin:2011ng,Schmookler:2019nvf}.
The nuclear cross-section for quasi-elastic scattering on a NN-SRC pair is
\begin{equation}
\begin{split}
\sigma_{\text{A}}(Q^2, x_{\text{B}}\sim 2)= A\frac{\hat{a}_2(A)}{2}
\sigma_{\text{2N}}(Q^2, x_{\text{B}}\sim 2),
\end{split}
\label{eq:QE-cross-section}
\end{equation}
where $\hat{a}_2(A)$ represents the probability of a nucleon being in a NN-SRC configuration,
$\sigma_{\text{2N}}(Q^2, x_{\text{B}}\sim 2)$ is the cross-section for the electron-SRC pair scattering,
and $A$ is the mass number of the nucleus.
Instead of measuring the absolute cross-section, the cross-section ratio is usually measured in experiment
to explore the nuclear effects and to reduce the systematic uncertainty. The SRC cross-section ratio $a_2(A)$ is defined as,
\begin{equation}
\begin{split}
a_2(A)=\frac{\sigma_{\text{A}}(Q^2, x_{\text{B}}\sim 2)/A}{\sigma_{\text{D}}(Q^2, x_{\text{B}}\sim 2)/2},
\end{split}
\label{eq:a2-def}
\end{equation}
where the deuteron (D) is the reference nucleus.
The properties of NN-SRC pairs in different nuclei are said to be universal
if the SRC ratio $a_2(A)$ has no kinematical dependences in the SRC region.
In this case Eq. (\ref{eq:QE-cross-section}) yields
\begin{equation}
\begin{split}
a_2(A)=\frac{\hat{a}_2(A)}{\hat{a}_2(D)},
\end{split}
\label{eq:a2-scaling}
\end{equation}
which is independent of the kinematic variables $Q^2$ and $x_{\rm B}$,
commonly called the ``scaling" \cite{Frankfurt:1993sp,Egiyan:2005hs,Fomin:2011ng,Schmookler:2019nvf}.
The ``scaling" of the ratio $a_2$ is observed at JLab using a high energy and high intensity electron beam.
Note that a high $Q^2$ ($Q^2>1$ GeV$^2$) is needed to reduce the final-state interactions,
and small $\nu$ is required to reduce the contamination by inelastic scattering \cite{Fomin:2011ng}.
In the review article on hard probes of NN-SRC \cite{Arrington:2011xs}, a moderate momentum transfer (1 GeV$^2<Q^2<$ 4 GeV$^2$)
is suggested to avoid the long-range two-body interactions and the contributions from $\Delta$-isobars.
The energy of the virtual photon is required to be much larger than the nucleon-nucleon potential.
The kinematical conditions on well separation of two-nucleon and three-nucleon SRCs
are also discussed in the literature \cite{Arrington:2011xs}, which provide a good guidance for the experiments.

The measured SRC ratio $a_2(A)$ connects to the relative number of SRC pairs
per nucleon, as the scattering at $x_{\rm B}\sim 2$ is dominantly attributed to the scattering between
the electron and the NN-SRC pairs. Therefore the number of NN-SRC pairs in a nucleus (of mass number $A$)
is expressed as the product of the measured SRC ratio $a_2$
and the number of SRC pairs in a deuteron, which is shown as in Eq. (\ref{eq:PairNumberAndA2}).
\begin{equation}
\begin{split}
a_2(A)=\frac{n^{\text{A}}_{\text{SRC}}/A}{n^{\text{d}}_{\text{SRC}}/2},\\
n^{\text{A}}_{\text{SRC}}=A \times a_2(A) \times \frac{n^{\text{d}}_{\text{SRC}}}{2}.
\end{split}
\label{eq:PairNumberAndA2}
\end{equation}
We make the approximation that both the nucleon mass in a NN-SRC pair ($m_{\text{SRC}}$)
and the nucleon mass in the mean-field ($m_{\text{MF}}$) are universal for all nuclei.
Then the nuclear mass is decomposed into two terms:
\begin{equation}
\begin{split}
M(A,Z)=2n^{\text{A}}_{\text{SRC}}m_{\text{SRC}}+(A-2n^{\text{A}}_{\text{SRC}})m_{\text{MF}}.
\end{split}
\label{eq:NuclearMassDecomposition}
\end{equation}
Here, the mass of the SRC nucleon is defined as the half of the mass of the SRC pair.
Combining Eqs. (\ref{eq:PairNumberAndA2}) and (\ref{eq:NuclearMassDecomposition}), we get the following decomposition of the nuclear mass per nucleon
\begin{equation}
\begin{split}
\frac{M(A,Z)}{A}=m_{\text{MF}}+a_2(A)n^{\text{d}}_{\text{SRC}}
(m_{\text{SRC}}-m_{\text{MF}}).
\end{split}
\label{eq:NuclearMassPerNucleon}
\end{equation}
For the deuteron, the SRC ratio $a_2(D)$ is one, hence we have a constraining condition,
\begin{equation}
\begin{split}
\frac{M^{\text{d}}}{2}=m_{\text{MF}}+n^{\text{d}}_{\text{SRC}}
(m_{\text{SRC}}-m_{\text{MF}}),
\end{split}
\label{eq:DMassAndSRC}
\end{equation}
for the nuclear mass decomposition Eq. (\ref{eq:NuclearMassPerNucleon}).
From Eq. (\ref{eq:DMassAndSRC}), we obtain $m_{\text{MF}}$ in terms of $n^{\text{d}}_{\text{SRC}}$
and $m_{\text{SRC}}$. Substituting $m_{\text{MF}}$ by the expression in terms of $n^{\text{d}}_{\text{SRC}}$
and $m_{\text{SRC}}$ into Eq. (\ref{eq:NuclearMassPerNucleon}), finally we find the nuclear mass per nucleon as a function
of $a_2(A)$, with two free parameters $n^{\text{d}}_{\text{SRC}}$ and $m_{\text{SRC}}$,
\begin{equation}
\begin{split}
\frac{M(A,Z)}{A}=\frac{M^{\text{d}}}{2(1-n^{\text{d}}_{\text{SRC}})}
-\frac{n^{\text{d}}_{\text{SRC}}}{1-n^{\text{d}}_{\text{SRC}}}m_{\text{SRC}} \\
+a_2(A)\left[ \frac{n^{\text{d}}_{\text{SRC}}}{1-n^{\text{d}}_{\text{SRC}}}m_{\text{SRC}}
- \frac{M^{\text{d}}n^{\text{d}}_{\text{SRC}}}{2(1-n^{\text{d}}_{\text{SRC}})}  \right].
\end{split}
\label{eq:NuclearMassAndA2}
\end{equation}

\begin{figure}[htp]
\centering
\includegraphics[width=0.48\textwidth]{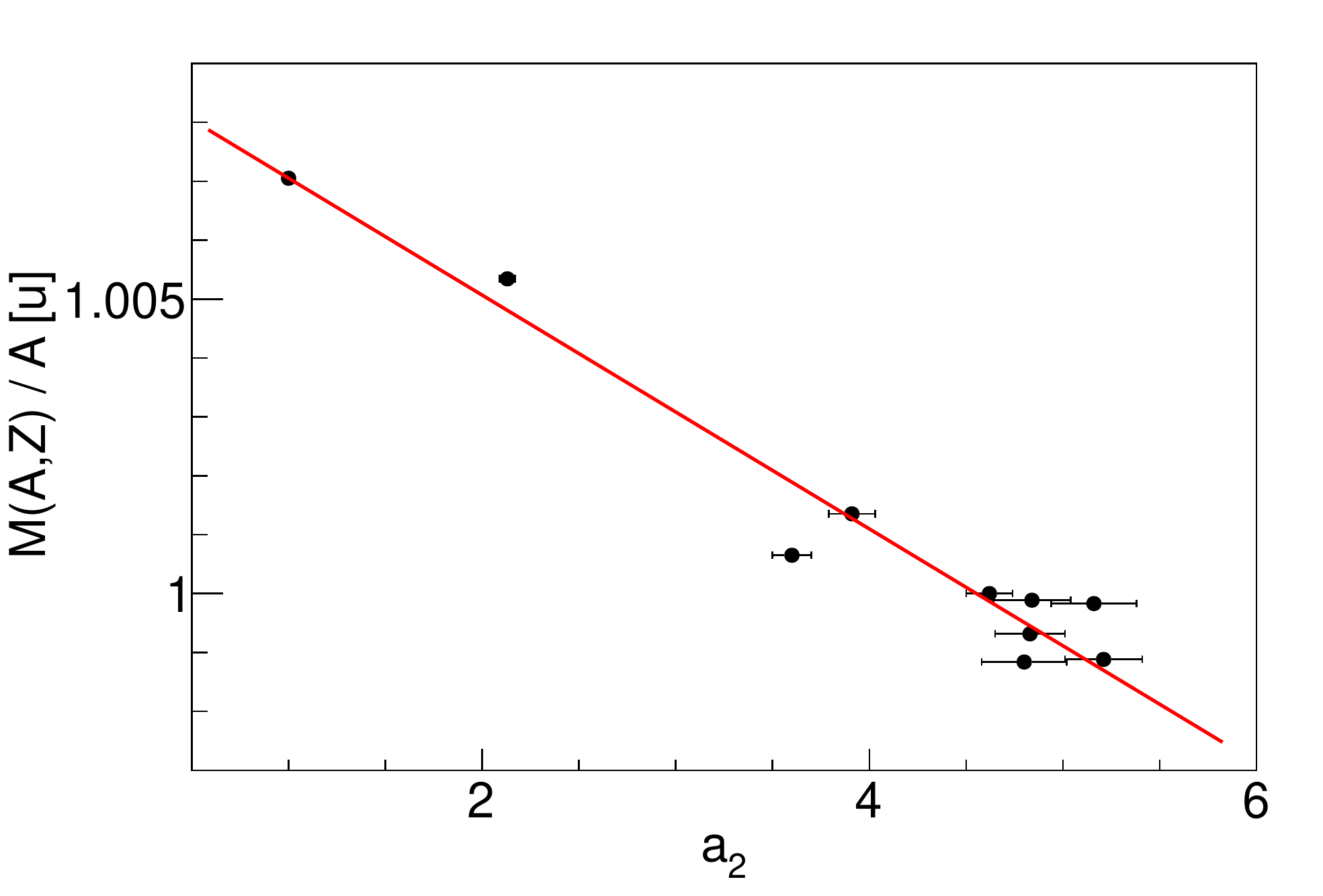}
\caption{
The correlation between the nuclear mass per nucleon and the NN-SRC ratio $a_2(A)$
\cite{Frankfurt:1993sp,Egiyan:2005hs,Fomin:2011ng,Schmookler:2019nvf,Hen:2012fm}.
The nuclear mass is reported in the atomic mass unit (1 u = 931.5 MeV).
}
\label{fig:a2-mass-fit}
\end{figure}

The nuclear mass as a function of $a_2$ is shown in Fig.~\ref{fig:a2-mass-fit}.
The nuclear masses are taken from the Refs. \cite{WangAME2016,HuangAME2016},
and the $a_2$ data are taken from the combination of the two analyses \cite{Schmookler:2019nvf,Hen:2012fm}
of different experimental measurements \cite{Frankfurt:1993sp,Egiyan:2005hs,Fomin:2011ng}.
A linear fit is performed to the mass-$a_2$ correlation using Eq. (\ref{eq:NuclearMassAndA2}).
All current data are more or less distributed around the fit.
The quality of the fit is $\chi^2/N=95/9=10.5$.
The nucleon mass in a NN-SRC pair is found to be $m_{\text{SRC}}=0.915\pm 0.019$ u $=852\pm 18$ MeV,
and the number of SRC pairs in a deuteron is $n^{\text{d}}_{\text{SRC}}=0.021\pm 0.005$.
Applying Eq. (\ref{eq:DMassAndSRC}), we obtain the mean-field nucleon mass $m_{\text{MF}}=1.009\pm 0.005$ u $=939.9\pm 5$ MeV.
Currently a theoretical calculation from the GCF method shows that
there are at least 20\% uncertainty when using the $a_2$ data
as the empirical SRC pair abundances \cite{Weiss:2020mns}.
Therefore the fit quality of the mass-$a_2$ correlation could be improved
if these model uncertainties larger than 20\% are included in the analysis.
The quality of the fit is reduced down to $\chi^2/N=1.6/9=0.2$
if the additional uncertainties of 20\% are considered in the analysis.

Regardless of the additional uncertainty of 20\% mentioned above for $a_2$ data interpretation,
we speculate that there are two other reasons that the quality of the fit is not good.
First, the mean-field nucleon mass is not universal for all nuclei.
Second, there are more terms in addition to the nuclear mass decomposition in Eq. (\ref{eq:NuclearMassDecomposition}),
such as the nucleon mass resulting from the multiple-nucleon short-range correlations (more than two nucleons).
The density of the light nucleus is quite different from the heavy one,
hence the mean-field nucleon mass of light nucleus possibly exhibits a sizeable difference.
If we exclude the data of $^3$He and $^4$He when performing the fit, we get an amazing quality of the fit
$\chi^2/N=9.1/7=1.3$. At the same time, the obtained nucleon mass in SRC and the number of NN-SRCs
are almost unchanged.

Thinking about two nucleon bags merging into one nucleon bag,
the volume of two nucleons is reduced by the volume of one nucleon bag.
Therefore for two completely overlapping nucleons, the vacuum energy is reduced
by the vacuum energy of one nucleon bag.
In the MIT bag model, in the relativistic-gas approximation, the volume of a stable ``bag"
is $V\approx B^{-3/4}$, where $B$ is the vacuum energy per unit volume.
The vacuum energy of one nucleon bag is then estimated to be $BV\approx B^{1/4}$.
Based on an analysis of the baryon ground state, $B^{1/4}$ is fit to be about 120 MeV \cite{Chodos:1974pn}.
As a result, the mass loss per nucleon in an NN-SRC pair is predicted
to be $B^{1/4}/2=60$ MeV, which will be denoted bag model-I in this work.
In this model, $B=(0.12)^4$ GeV$^4$ $=0.027$ GeV fm$^{-3}$ also equals the external pressure at the ``bag" boundary,
which is actually close to the preliminary result of the pressure inside a proton $p(r=0.85~\text{fm})\sim 0.04$ GeV fm$^{-3}$
in a recent analysis of generalized parton distribution functions from DVCS experiments \cite{Burkert:2018bqq}.
The equilibrium condition and the relativistic-gas approximation in the bag model gives,
\begin{equation}
\begin{split}
p=\frac{1}{3}\frac{E_r}{V}=B,\\
E=E_r+BV=4BV,
\end{split}
\label{eq:BagModel-E}
\end{equation}
where $E_r$ is the internal energy carried by quarks and gluons, and $E$ is the total energy \cite{Chodos:1974je}.
Therefore based on the equilibrium condition (Eq. (\ref{eq:BagModel-E})), the vacuum energy of one nucleon
is $BV=E/4=m_{\rm N}/4$. This gives a mass loss per nucleon in a NN-SRC pair
as $m_{\rm N}/8=117$ MeV, labeled as bag model-II in this work.

In QCD theory, the trace anomaly contribution $M_{\rm a}$ to the nucleon mass
is analogous to the vacuum energy in the MIT bag model, i.e. the vacuum energy of one nucleon bag.
The energy loss (mass deficit) of NN-SRC equals the trace anomaly $M_{\rm a}$,
for the volume reduction of the extremely strong NN-SRC configuration
is the volume of one nucleon, compared to that of two free nucleons.
The vacuum energy loss per nucleon is therefore
half of the trace anomaly contribution to the nucleon mass.
According to a LQCD calculation \cite{Yang:2018nqn} on the QCD trace anomaly,
this corresponds to a mass loss per paired nucleon of $M_{\rm a}/2=108$ MeV.
Based on this assumption concerning the trace anomaly, the mass loss of a nucleon in a 3N-SRC
is $2M_{\rm a}/3$, and the mass loss of the nucleon in a 4N-SRC is $3M_{\rm a}/4$, and so on.

\begin{table}[h]
	\caption{List of the mass deficits of a nucleon in NN-SRC
             from this work and some models. See the main text for detailed explanations.  }
		\begin{tabular}{cc}
        \hline\hline
         Method      &  $m_{\text{N}}-m_{\text{SRC}}$   \\
        \hline
         This analysis & 86 $\pm$ 18 MeV   \\
         Bag model-I  \cite{Chodos:1974pn}   & $B^{1/4}/2=60$ MeV   \\
         Bag model-II \cite{Chodos:1974je}   & $m_{\rm N}/8=117$ MeV   \\
         Trace anomaly \cite{Yang:2018nqn}   & $M_{\rm a}/2=$108 MeV   \\
         QMC model \cite{Saito:2005rv}       & 137 $\sim$ 163 MeV   \\
        \hline\hline
		\end{tabular}
\label{Tab:mass-defect-list}
\end{table}

Although the mesons are not the fundamental degrees of freedom in QCD,
they give a good approximation to the nuclear force at low energy.
The quark-meson coupling (QMC) model, a mean-field description of non-overlapping
nucleon bags bound by the self-consistent exchange of mesons, predicts an effective nucleon
mass of $775\sim 801$ MeV in nuclear matter at saturation density,
with the structure effects of both the nucleon and the mesons considered \cite{Saito:2005rv}.
Hence the effective nucleon mass in the nuclear environment of high density puts
some new constraints on the parameters of QMC model, such as the various meson-quark couplings
and the nucleon form factors in nuclear matter.

To sum up, table \ref{Tab:mass-defect-list} lists the values of the mass deficit of a NN-SRC nucleon
from this analysis, and some models.
In our analysis of $a_2$ data, each nucleon in a NN-SRC pair loses a mass of $86\pm 18$ MeV.
We find that our extracted value of the nucleon mass loss
in a NN-SRC state is consistent with the QCD trace anomaly part of nucleon mass and
the MIT bag model predictions.

The number of SRC pairs in a deuteron is an interesting quantity, which tells us the probability of the nucleon
being in a short-distance configuration. From our fit, the nucleons in a deuteron have about 2.1\%
probability of forming a SRC pair.
The nuclear tensor force dominated at medium distance plays a crucial role in the formation
of pn-SRC pairs. Therefore the exact probability of pn-SRC pairing in deuteron
gives some significant constraints on the details of the tensor force.
This could be studied with the future numerical calculations based on $ab~initio$ method,
or the effective method such as the GCF model.
Using Eq. (\ref{eq:PairNumberAndA2}), we calculate the number of pn-SRC pairs in $^{12}$C
to be $n_{\text{SRC}}^{\text{C}}=0.58\pm 0.14$, with the obtained $n^{\text{d}}_{\text{SRC}}=0.021\pm 0.005$.
Based on our analysis, 9.7\% $\pm$ 2.4\% of the nucleons in $^{12}$C are in the NN-SRC state,
which is smaller than the previous estimations \cite{Subedi:2008zz,Egiyan:2005hs}.

In summary, we have obtained the nucleon mass in NN-SRC, which shows that
the nucleon in a SRC pair has a large mass deficit around 86 MeV, in agreement with
a QCD decomposition of the nucleon mass and the MIT bag model predictions.
One feature of NN-SRC is the high momentum of the nucleon inside the pair,
which actually requires a far off-shell SRC nucleon.
Moreover, we also find the number of NN-SRC pairs in a deuteron to be $0.021\pm 0.005$,
from the correlation between the nuclear mass and the SRC ratio $a_2(A)$.
Our analysis suggests that about 9.7\% $\pm$ 2.4\% of nucleons
are in pn-SRC configurations for the commonly studied nucleus $^{12}$C.

\begin{acknowledgments}
We thank Dr. Jarah Evslin for reading the manuscript and the help of improving the English writing.
This work is supported by the Strategic Priority Research Program of Chinese Academy of Sciences under the Grant NO. XDB34030301,
and the National Science Foundation China under the Grant NO. 11305007.
\end{acknowledgments}

\bibliographystyle{apsrev4-1}
\bibliography{refs}

\end{document}